\documentclass[aps,pre,reprint,superscriptaddress,10pt]{revtex4-2}
\usepackage{amsmath}
\usepackage{graphicx}
\usepackage{dcolumn}
\usepackage{bm}
\usepackage{hyperref}
\usepackage{courier}
\usepackage{braket}
\usepackage{float}
\usepackage[usenames,dvipsnames]{xcolor}
\hypersetup{
  colorlinks   = true,	
  urlcolor     = Fuchsia,	
  linkcolor    = Fuchsia,	
  citecolor   = DarkOrchid	
}
\newcommand{\cc}{\color{black}}
\begin{document}
\newcommand{\ve}{\varepsilon}
\title{Kerr nonlinearity hinders symmetry-breaking states of coupled quantum oscillators}

\author{Biswabibek Bandyopadhyay}
\affiliation{Chaos and Complex Systems Research Laboratory, Department of Physics, University of Burdwan, Burdwan 713
  104, West Bengal, India}
\author{Tanmoy Banerjee}
\email[]{tbanerjee@phys.buruniv.ac.in}
\thanks{he/his/him}
\altaffiliation{}
\affiliation{Chaos and Complex Systems Research Laboratory, Department of Physics, University of Burdwan, Burdwan 713
  104, West Bengal, India}
\date{\today}

\begin{abstract}
We study the effect of Kerr anharmonicity on the symmetry breaking phenomena of coupled quantum oscillators. Two types of symmetry-breaking processes are studied, namely the inhomogeneous steady state (or quantum oscillation death state) and quantum chimera state. Remarkably, it is found that Kerr nonlinearity hinders the process of symmetry-breaking in both the cases. We establish our results using direct simulation of quantum master equation and analysis of the stochastic semiclassical model. Interestingly, in the case of quantum oscillation death, an increase in the strength of Kerr nonlinearity tends to favor the symmetry and at the same time decreases the degree of quantum mechanical entanglement. This study presents a useful mean to control and engineer symmetry-breaking states for quantum technology.
\end{abstract}

\maketitle

\section{Introduction}
Spontaneous symmetry breaking is a fundamental phenomenon in nature and has been widely studied in the field of physics, chemistry, and biology \cite{moter}. In the context of coupled oscillators, symmetry-breaking has two broad manifestations, namely the oscillation death state \cite{kosprl,dvprep,tanpre1} and chimera state \cite{schoell_rev,annabook}. Oscillation death (OD) is the manifestation of inhomogeneous steady states that appear due to coupling dependent symmetry breaking \cite{kosprep,scholl-epl,tanpre2}. On the other hand, spontaneous symmetry breaking  in a network of identical oscillators gives rise to chimeras where coherent and incoherent groups of oscillators coexist \cite{KuBa02,scholl_CD,tanCD}. {\cc Therefore, in the OD state, the breaking of symmetry refers to breaking of a rotational invariance in phase space, while in chimeras, it refers to breaking of a symmetry among the oscillators}. Apart from their fundamental interest, OD and chimeras are generally linked to unihemispheric sleep of certain aquatic animals and migratory birds, cellular differentiation, and species persistence in ecology \cite{PaAb15,qstr,qstr2,bandutta,lr16}.

Looking at the importance of the symmetry breaking states, they have been searched and discovered recently in the quantum regime. Bastidas \textit{et al.} \cite{qchm} reported the quantum signature of chimera state where the notion of synchrony and asynchrony were determined according to the direction of squeezing in the phase space. The quantum oscillation death (QOD) and its diverse manifestations has been reported recently by the present authors \cite{qmod,qrev,qturing} that is manifested in the phase space in the form of a bimodal Wigner function. These studies are motivated by the continuous pursuit of understanding emergent behaviors of coupled oscillators in the quantum regime. It has been found that the well known collective dynamics behave in a different manner in the quantum domain due to the inherent {\cc quantum mechanical constraints stem from Heisenberg uncertainty principle and discreteness of energy levels}. For example, the well known synchronization scenario shows a counter intuitive behavior in the quantum regime \cite{lee_prl,brud_prl1,lee_pre_1,brud-ann15,squeezing,chia,expt1,expt2}, where no true phase-locking is possible due to the inherent quantum noise.  

In quantum systems and processes, Kerr nonlinearity is found to be present and relevant \cite{knightbook}. In the context of emergent dynamics of coupled quantum oscillators, the effect of Kerr nonlinearity was explored in the quantum synchronization scenario and quantum amplitude death state. Amitai \textit{et. al.} \cite{qad2} reported that Kerr nonlinearity promotes amplitude death. They further found multiple resonances in the mean phonon number and thereby revival of oscillation. Ref.~\cite{brud-poch} showed the genuine quantum signature of synchronization of driven anharmonic oscillator that shows nonclassical states under certain conditions. L\"{o}rch \textit{et al.} \cite{blockade} discovered that energy quantization deteriorates synchronization in Kerr-type limit cycle oscillators. However, surprisingly, the {\it effect of Kerr nonlinearity on the symmetry-breaking states} is yet to be studied.

In this paper, for the first time, we explore the effect of Kerr nonlinearity on the quantum symmetry-breaking states. We consider two quantum van der Pol oscillators under conjugate coupling and reveal that the Kerr anharmonicity hinders the symmetry breaking states and prefers the symmetry preserving states. We demonstrate our results in both type of symmetry-breaking states, i.e., quantum oscillation death state and quantum chimera state. Our observation is supported by the direct simulation of quantum master equation, analysis of the noisy classical model, and the degree of quantum mechanical entanglement. 


\section{Effect of Kerr nonlinearity on symmetry breaking death state}

\subsection{Mathematical model}
\label{sec:vdp}

The quantum master equation in density matrix $\rho$ of a quantum van der Pol (vdP) oscillator in the presence of Kerr nonlinearity reads (setting $\hbar=1$) \cite{lee_prl,brud_prl1,blockade}
\begin{equation}
\label{master-single}
\dot{\rho}=-i[H,\rho]+k_1\mathcal{D}[{a}^\dag](\rho)+k_2\mathcal{D}[{a}^2](\rho),
\end{equation}
where $H=\omega ({a}^\dag a)+K({a}^\dag a)^2$. Here $\omega$ is the natural frequency of the vdP oscillator, and $K$ is the Kerr nonlinearity parameter. $\mathcal{D}[\hat{L}](\rho)$ is the Lindblad dissipator having the form $\mathcal{D}[\hat{L}](\rho)=\hat{L}\rho \hat{L}^\dag-\frac{1}{2}\{\hat{L}^\dag \hat{L},\rho \}$, where $\hat{L}$ is an operator. $a$ and ${a}^\dag$ are the bosonic anihilation and creation operators, respectively. $k_1$ governs the rate of linear pumping and $k_2$ governs the rate of nonlinear damping. For $K=0$, Eq.~\eqref{master-single} gives the conventional quantum vdP oscillator that exhibits a limit cycle oscillation. In the presence of Kerr nonlinearity ($K>0$) the oscillator becomes anharmonic; It has been shown in Ref.~\cite{brud-poch} that even for $K>0$ the limit cycle is preserved without any phase preference.  

Next, we consider two quantum vdP oscillators with Kerr nonlinearity coupled via conjugate coupling. Ref.~\cite{qturing} reported that quantum vdP oscillators under this coupling show symmetry-breaking steady state, i.e., the quantum oscillation death state. Here the quantum master equation of conjugately coupled quantum vdP oscillators in the presence of Kerr nonlinearity reads
\begin{equation}
\label{master}
\begin{split}
\dot{\rho}&=-i[H_c,\rho]+k_1\sum_{j=1}^{2}\mathcal{D}[{a_j}^\dag](\rho)\\
&+k_2\sum_{j=1}^{2}\mathcal{D}[{a_j}^2](\rho)+\ve\sum_{j=1}^{2}\mathcal{D}[a_j](\rho),
\end{split}
\end{equation}
where the Hamiltonian of the coupled system is given by: $H_c=\omega ({a_1}^\dag a_1+{a_2}^\dag a_2)+K({a_1}^\dag a_1)^2+K({a_2}^\dag a_2)^2+\frac{\ve}{2}({a_1}^\dag a_2+{a_2}^\dag a_1)-\frac{\ve}{2}({a_1}^\dag {a_2}^\dag+a_1a_2)-\frac{i\ve}{4}({{a_1}^\dag}^2+{{a_2}^\dag}^2-{a_1}^2-{a_2}^2)$. $\varepsilon$ is the coupling strength ($\varepsilon \ge 0$).  In the following we will explore the effect of Kerr nonlinearity ($K$) on the system dynamics.

\subsection{Numerical results: Quantum model}\label{sec:results}

We directly solve the master equation \eqref{master} using \texttt{QuTiP} \cite{qutip}. {\cc Without any loss of generality, throughout the paper we consider the natural frequency of the oscillators $\omega=2$}. To visualize and understand the system dynamics we employ the Wigner function representation in phase space \cite{wigner}. {\cc As the oscillators are identical, they will show identical dynamics, therefore, it is sufficient to track one of the oscillators. Further, to distinguish between QAD and QOD, we consider only the weak quantum regime where $k_2<k_1$ \cite{qturing}}.
\begin{figure}
\includegraphics[width=.48\textwidth]{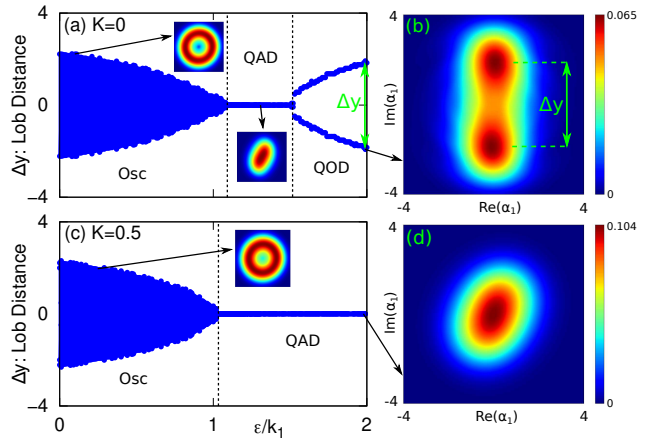}
\caption{\textbf{Numerical results of quantum model}: (a) Quantum bifurcation diagram showing the variation of $y$-projection of the distance between the local maximum value of Wigner distribution in phase space with increasing $\varepsilon/k_1$ at Kerr parameter $K=0$. Insets show the Wigner function of quantum limit cycle at $\varepsilon/k_1=0.1$, and QAD at $\varepsilon/k_1=1.4$. (b) The Wigner distribution function in phase space for a representative point ($\varepsilon/k_1=1.99$) of (a). (c) Quantum bifurcation diagram for Kerr parameter $K=0.5$. {\cc Note that the QOD state does not appear in the same scale of Fig.~\ref{kerr_quantum}(a); instead, the boundary to the QOD region is moved towards larger $\epsilon/k_1$ (see Fig.~\ref{f:2par})}. The inset shows quantum limit cycle in phase space ($\varepsilon/k_1=0.25$). (d) The Wigner function distribution in phase space for a representative point ($\varepsilon/k_1=1.99$) of (c). Other parameters are $\omega=2$ and $(k_1, k_2)=(1, 0.2)$.}
\label{kerr_quantum}
\end{figure}
To draw the quantum bifurcation scenario we define a variable $\Delta y$, which is the $y$-projection of the distance between local maxima of the Wigner function. Figure~\ref{kerr_quantum}(a) represents the quantum bifurcation diagram showing the variation of $\Delta y$ with the coupling strength $\varepsilon/k_1$ in the absence of Kerr anharmonicity (i.e., $K=0$). For lower values of coupling strength the coupled system shows quantum limit cycle; the Wigner function of the limit cycle is shown in the inset of Fig.~\ref{kerr_quantum}(a) at $\varepsilon/k_1=0.1$, 
{\cc which is a ring shaped structure. For a particular value of coupling strength ($\varepsilon/k_1$), by plotting the $y$-coordinates of all the points of local maxima on the ring we get a vertical line whose length is equal to the diameter of the ring. With increasing coupling strength, the size of the ring decreases and so does the diameter of the ring. That is why we get shorter vertical lines as we move towards higher $\varepsilon/k_1$. The stack of these vertical lines creates the shape of blue filled area in Fig.~\ref{kerr_quantum}(a, c) representing quantum limit cycles.}
Beyond a value of $\varepsilon/k_1$ quantum amplitude death (QAD) state appears that is indicated by a single line at $\Delta y=0$. The Wigner function in the QAD state is shown in the inset of Fig.~\ref{kerr_quantum}(a) for $\varepsilon/k_1=1.4$: it shows a squeezed amplitude death state. However, unlike classical case a complete suppression of oscillations is prevented by the inherent quantum noise present in the system. Further increment in coupling strength gives rise to a symmetry-breaking transition from quantum amplitude death (QAD) state to quantum oscillation death (QOD) state. In the QOD state we get two branches in the bifurcation diagram with $\Delta y \neq 0$. Fig.~\ref{kerr_quantum}(b) shows the Wigner function representation in phase space at $\varepsilon/k_1=1.99$ that shows a bimodal structure, which is the manifestation of QOD in the phase space.

\begin{figure}
\includegraphics[width=.48\textwidth]{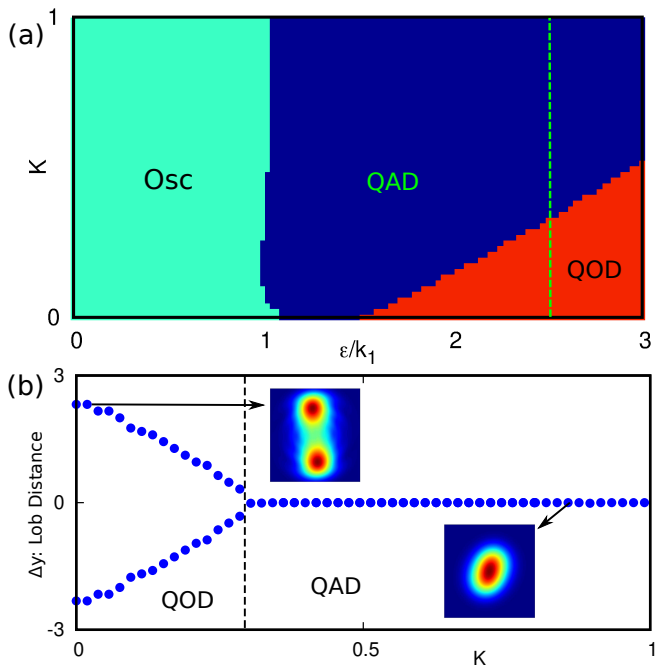}
\caption{(a) Two parameter phase diagram in the $\varepsilon/k_1-K$ space delineating the zone of quantum limit cycle (Osc), quantum amplitude death (QAD) and quantum oscillation death (QOD) states. (b) Bifurcation diagram with $K$ along the vertical dashed line of (a), i.e., $\varepsilon/k_1=2.5$. insets show the Wigner distribution function of QOD ($K=0.02$) and QAD  ($K=0.85$). Other parameters are $\omega=2$ and ($k_1,k_2$)=($1,0.2$).}
\label{f:2par}
\end{figure}

Now we demonstrate the effect of Kerr anharmonicity in the system by choosing an exemplary value $K=0.5$. Fig.~\ref{kerr_quantum}(c) represents the quantum bifurcation diagram of the anharmonic system. Here we observe that the QOD state does not appear {\cc in the same scale of Fig.~\ref{kerr_quantum}(a)}, instead the QAD region (single branch, $\Delta y=0$) persists all the way. {\cc In fact, the boundary to the QOD region is moved towards larger $\epsilon/k_1$ (see Fig.~\ref{f:2par})}. Plotting the Wigner function in the phase space at the same value of coupling strength $\varepsilon/k_1=1.99$, we get a unimodal structure (Fig.~\ref{kerr_quantum}(d)) indicating QAD, i.e., the Kerr parameter is conducive for the symmetric state rather than the symmetry-breaking state.  

For a complete understanding of the scenario, we explore the effect of $K$ in the $\varepsilon/k_1-K$ parameter space. Fig.~\ref{f:2par}(a) delineates the zone of quantum limit cycle (Osc), quantum amplitude death (QAD) and quantum oscillation death (QOD) states in the two parameter phase diagram for ($k_1,k_2$)=($1,0.2$). It clearly shows that if we start from a symmetry-breaking state (i.e., QOD state), an increase in the Kerr parameter ($K$) eventually destroys that state and brings the coupled system to a symmetric homogeneous state, namely the QAD state. 
Next, we draw a bifurcation diagram of the lobe distance ($\Delta y$) with $K$ at $\varepsilon/k_1=2.5$ [along the vertical line of Fig.~\ref{f:2par}(a)]. Fig.~\ref{f:2par}(b) depicts that with an increasing $K$ the symmetry-breaking QOD state eventually converts into a QAD state. Insets of Fig.~\ref{f:2par}(b) also show the Wigner distribution of the corresponding states. {\cc Note that, in the QAD state, a slight phase preference still exists, which does not violate our definition of QAD, which is based on the unimodal Wigner function.}
Therefore, we infer that the Kerr anhamonicity hinders the symmetry-breaking state and  prefers symmetry preserving oscillation quenching state. 

\subsection{Comparison to noisy-classical model}\label{sec:noisyclassical} 
We verify our results analytically in the semiclassical regime. In the semiclassical limit, linear pumping dominates over the nonlinear damping (i.e., $k_1>k_2$) and one can approximate $\langle a\rangle \equiv \alpha$, and starting from the master equation \eqref{master}, using the relation $\dot{\braket{a}}=\mbox{Tr}(\dot \rho a)$, one arrives at the following classical amplitude equation:
\begin{equation}
\label{amp}
\begin{split}
\dot{\alpha_j}&=-i\left(\omega+2K|\alpha_j|^2\right)\alpha_j+(\frac{k_1}{2}-k_2|\alpha_j|^2)\alpha_j \\
&-\frac{\ve}{2}\left[(\alpha_j+{\alpha_j}^*)+i(\alpha_{j'}-{\alpha_{j'}}^*)\right],
\end{split}
\end{equation}
{\cc where $(j,j') = (1,2)$ and $(2,1)$}. We compare the results obtained from the quantum master equation of the previous section with the corresponding noisy classical model (or semiclassical model). In the noisy classical model the classical dynamics is considered in the presence of a finite amount of noise whose intensity is equal to that of the quantum noise. To evaluate the quantum noise intensity, a stochastic differential equation is derived from the quantum master equation following Ref.~\cite{qad1}. 
For this, the quantum master equation \eqref{master} is represented in the phase space using a partial differential equation of Wigner distribution function ($W(\bm{\alpha})$) \cite{carmichael}: 
\begin{equation}
\label{diff_eqn_w}
\begin{split}
\partial_t{W(\bm{\alpha})}&=\sum_{j=1}^2\left[-\left(\frac{\partial}{\partial \alpha _j}\mu _{\alpha _j}+c.c.\right) \right. \\
&+ \left. \frac{1}{2}\left(\frac{\partial ^2}{\partial \alpha _j \partial {\alpha_j}^*}D_{\alpha _j {\alpha_j}^*}+\frac{\partial ^2}{\partial \alpha _j \partial {\alpha_{j'}}^*}D_{\alpha _j {\alpha_{j'}}^*} \right) \right. \\
&+ \left. \frac{k_2}{4}\left(\frac{\partial ^3}{\partial {\alpha_j}^* \partial {\alpha_j}^2}\alpha _j+c.c\right) \right]W(\bm{\alpha}),
\end{split}
\end{equation}
where the elements of the drift vector ($\bm{\mu}$) are: $\mu _{\alpha _j}=\left[-i\left(\omega+2K|\alpha _j|^2\right)+\frac{k_1}{2}-k_2(|\alpha _j|^2-1)-\frac{\ve}{2}\right]\alpha _j-\frac{\ve}{2}{\alpha_{j}}^*-\frac{i\ve}{2}{\alpha_{j'}}+\frac{i\ve}{2}{\alpha_{j'}}^*,$
and the elements of the diffusion matrix $\bm{D}$ are:
$D_{\alpha _j {\alpha_j}^*}=k_1+2k_2(2|\alpha _j|^2-1)+\ve, D_{\alpha _j {\alpha_{j'}}^*}=0$, with $j=1,2$, $j'=1,2$ and $j\neq j'$. In the weak nonlinear regime ($k_2\ll k_1$), {\cc the third term inside the bracket of Eq.~\eqref{diff_eqn_w} having coefficient $\frac{k_2}{4}$ can be neglected}, and thus Eq.~\eqref{diff_eqn_w} reduces to the Fokker-Planck equation, which is given by 
\begin{equation}
\label{fp}
\begin{split}
\partial_t{W}(\textbf{X})&=\sum_{j=1}^2\left[-\left(\frac{\partial}{\partial x_j}\mu _{x_j}+\frac{\partial}{\partial y_j}\mu _{y_j}\right) \right) \\
&+ \left. \frac{1}{2}\left(\frac{\partial ^2}{\partial x_j \partial x_j}D_{x_j x_j}+\frac{\partial ^2}{\partial y_j \partial y_j}D_{y_j y_j} \right. \right. \\
&+ \left. \left. \frac{\partial ^2}{\partial x_j \partial x_{j'}}D_{x_j x_{j'}}+\frac{\partial ^2}{\partial y_j \partial y_{j'}}D_{y_j y_{j'}}  \right) \right]W(\textbf{X}),
\end{split}
\end{equation}
where $\textbf{X}=(x_1, y_1, x_2, y_2)$. The elements of drift vector are,
\begin{subequations}
\label{drift}
\begin{align}
\begin{split}
\mu _{x_j}&=\left(\omega+2K({x_j}^2+{y_j}^2)\right)y_j\\
&+ \left[\frac{k_1}{2}-k_2({x_j}^2+{y_j}^2-1)-\ve\right]x_j+ \ve y_{j'},
\end{split} \\
\begin{split}
\mu _{y_j}&=-\left(\omega+2K({x_j}^2+{y_j}^2)\right)x_j\\
&+ \left[\frac{k_1}{2}-k_2({x_j}^2+{y_j}^2-1)\right]y_j.
\end{split}
\end{align}
\end{subequations}
The diffusion matrix has the following form,
\begin{equation}\label{diffusion_mat} 
{\bm{D}}=\frac{1}{2}\left(\begin{array}{cccc} \nu _1 & 0 & 0 & 0 \\
0 & \nu _1 & 0 & 0\\
0 & 0 & \nu _2 & 0 \\
 0 & 0 & 0 & \nu _2  \end{array}\right).
\end{equation}
where $\nu _j=\frac{k_1}{2}+k_2[2({x_j}^2+{y_j}^2)-1]+\frac{\ve}{2}$.
From Eq.~\eqref{fp}, the following stochastic differential equation can be derived,
\begin{equation}\label{sde}
d\textbf{X}=\bm{\mu}dt+\bm{\sigma} d\textbf{W}_t,
\end{equation}
where $\bm{\sigma}$ is the noise strength and $d\textbf{W}_t$ is the Wiener increment. As the diffusion matrix $\bm{D}$ (given in Eq.~\eqref{diffusion_mat}) is diagonal, we can analytically derive $\bm{\sigma}$ from it as $\bm{\sigma}=\sqrt{\bm{D}}$.

By solving the stochastic differential equation (Eq.~\eqref{sde}) (using JiTCSDE module in Python \cite{jitcode}), we compute the behaviour of the noisy-classical system in phase space, starting from random initial conditions.

\begin{figure}
\vspace{0.4cm}
\includegraphics[width=.48\textwidth]{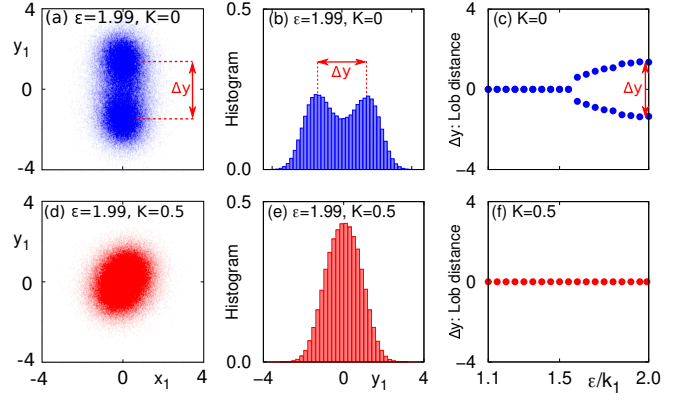}
\caption{\textbf{Analytical results of noisy-classical model}: Upper row (a-c) corresponds to system without anharmonicity, $K=0$. (a) Phase space plot and (b) histogram of $y_1$ variable at $\varepsilon/k_1=1.99$, showing bimodal shape (noisy-classical OD). (c) The bifurcation diagram showing the variation of $\Delta y$ (The $y$-projection of distance between two local maxima). Lower row (d-f) corresponds to the system with anharmonicity ($K=0.5$). At the same value of coupling strength ($\varepsilon/k_1=1.99$) (d) phase space plot and (e) Histogram of $y_1$ variable showing unimodal shape (noisy-classical AD). (f) Bifurcation diagram of $\Delta y$ showing AD. Other parameters are $\omega=2$ and ($k_1,k_2$)=($1,0.2$).}
\label{kerr_semicl}
\end{figure}

The analytical results of the noisy-classical model are demonstrated in Fig.~\ref{kerr_semicl}. The upper row Fig.~\ref{kerr_semicl} (a-c) corresponds to the system without Kerr anharmonicity ($K=0$) and the lower row Fig.~\ref{kerr_semicl}(d-f) corresponds to the system with Kerr anharmonicity ($K=0.5$). Fig.~\ref{kerr_semicl}(a) shows a bimodal distribution in phase space at $\varepsilon/k_1=1.99$ and $K=0$. This is in well agreement with the phase space representation of Wigner function at the same parameter value [shown in Fig.\ref{kerr_quantum}(b)]. Fig.~\ref{kerr_semicl}(b) represents the corresponding histogram of $y$-variable. This helps us to clearly visualize the bimodal structure of Fig.~\ref{kerr_semicl}(a), showing as double hump. Fig.\ref{kerr_semicl}(c) is the bifurcation diagram of the noisy-classical system in the absence of Kerr anharmonicity. 

In the presence of Kerr nonlinearity, the phase space diagram and histogram of the $y$-variable of the noisy-classical system are shown in Fig.~\ref{kerr_semicl}(d) and (e), respectively for $K=0.5$. The central unimodal lob shape of phase space diagram indicates the homogeneous steady state. This unimodal phase space representation is in accordance with the phase space representation of Wigner function for quantum system as shown in Fig.~\ref{kerr_quantum}(d) for the QAD. The histogram with a single hump at $y_1=0$ supports the occurrence of homogeneous steady state. In Fig.~\ref{kerr_semicl}(f) it is clearly visible that the symmetry breaking transition is totally swept out from the same parametric zone as chosen in Fig.~\ref{kerr_semicl}(c). Therefore, the results of noisy-classical model establishes that Kerr anharmonicity indeed hinders the symmetry-breaking state and induces symmetric state. 
\begin{figure}
\includegraphics[width=.48\textwidth]{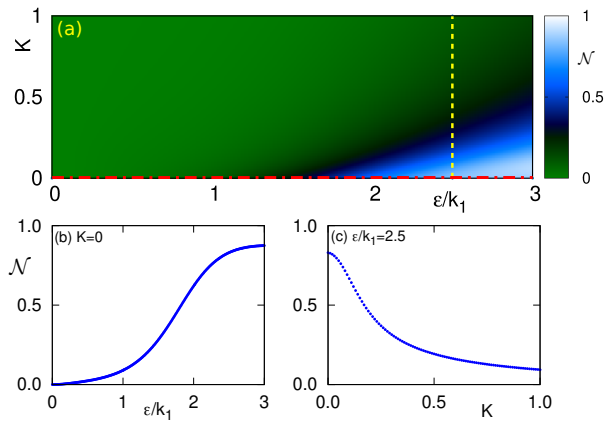}
\caption{\textbf{Effect of Kerr nonlinearity on quantum entanglement}: (a) Variation of negativity ($\mathcal{N}$) in $\varepsilon/k_1$ vs $K$ space. (b) Plot of negativity ($\mathcal{N}$) with parameter $\varepsilon/k_1$ corresponding to the horizontal line in (a) at $K=0$. (c) Plot of negativity with Kerr parameter ($K$) corresponding to the vertical line in (a) at $\varepsilon/k_1=2.5$. Other parameters are $\omega=2$ and ($k_1,k_2$)=($1,0.2$).}
\label{negativity}
\end{figure}
\begin{figure*}
\includegraphics[width=0.8\textwidth]{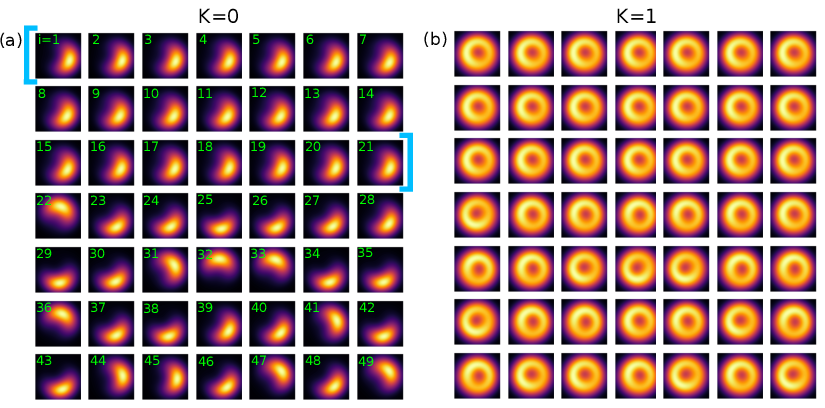}
\caption{\textbf{Effect of Kerr nonlinearity on quantum chimera state}: \textbf{(a) No Kerr ($K=0$):} After a short time evolution $\Delta t=1$, the initial coherent states evolve to squeezed states---shown using the Husimi Q-function in the phase space. The directions of squeezing are same for $i=1$ to $i=21$ oscillators forming the coherent group (indicated within $[..]$) and random for $i=22$ to $i=49$ oscillators forming the incoherent group: quantum signature of chimera. \textbf{(b) With Kerr ($K=1$):} The signature of quantum chimera is destroyed as all the oscillators show ring shape symmetric Husimi Q-function in the phase space. The distinction of coherent-incoherent group no longer exists. Other parameters are $k_1=1$, $k_2=0.2$, $d=10$ and $V=1.2$. For all the plots $x$-axis represents $Re(\alpha_i)$ ranging from $-4$ to $4$ and $y$-axis represents $Im(\alpha_i)$ also ranging from $-4$ to $4$.}
\label{kerrchm}
\end{figure*}
\subsection{Effect of Kerr nonlinearity on entanglement}
\label{sec:negativity}
Ref.~\cite{kato} established that symmetry-breaking states support quantum mechanical entanglement.
Here we investigate how the Kerr nonlinearity affects the phenomenon of quantum entanglement. For this we employ the negativity parameter ($\mathcal{N}$), which is a measure of quantum entanglement \cite{ent1,ent2}.
By computing negativity in our system we can predict the entanglement between two oscillators. Negativity is computed from the formula $\mathcal{N}=(||\rho^{\Gamma_1}||_1-1)/2$, where $\rho^{\Gamma_1}$ is the partial transpose of the density matrix $\rho$ of the system with respect to first oscillator and $||X||_1=Tr|X|=Tr\sqrt{X^{\dag}X}$ (here $X=\rho^{\Gamma_1}$). $\mathcal{N}=0$ indicates the absence of entanglement, as $\mathcal{N}$ increases entanglement enhances. 

Fig.~\ref{negativity}(a) shows the two parameter plot of negativity ($\mathcal{N}$) in the $\varepsilon/k_1$ vs $K$ space. Note the enhancement of $\mathcal{N}$ for the higher $\varepsilon$ and lower $K$ region, which qualitatively matches with the the QOD zone of Fig.~\ref{f:2par} (a). Two representative single parameter plots are shown in Figs.~\ref{negativity}(b) and (c) [along, the horizontal and vertical lines of Fig.~\ref{negativity}(a), respectively]. In Fig.~\ref{negativity}(b) (at $K=0$) we observe that in the uncoupled state ($\varepsilon/k_1=0$) the negativity is zero, indicating no entanglement of two quantum oscillators. As the coupling strength increases the entanglement becomes stronger with increment of the value of negativity. Comparing this with Fig.~\ref{kerr_quantum}(a) we can say that in QAD state the two coupled quantum oscillators are weakly entangled and the entanglement is  stronger in the QOD state.
From Fig.~\ref{negativity}(c) it is clear that, for $\varepsilon/k_1=2.5$ starting from $K=0$ (QOD state) as we increase the Kerr parameter, negativity decreases. Comparing this with Fig.~\ref{f:2par}(b) again confirms that the transition from QOD to QAD is indeed marked by a decreasing $\mathcal{N}$. Therefore, through the negativity parameter we also verify that Kerr nonlinearity hinders entanglement and therefore does not favor the symmetry-breaking state.

\section{Effect of Kerr nonlinearity on quantum chimera state}
\label{ch}
Quantum signature of chimera state was reported by Bastidas \textit{et al.} \cite{qchm}. They considered a nonlocally coupled ring of $N$ quantum Stuart-Landau oscillators, which obey the following master equation,
\begin{equation}
\label{qchm_master}
\dot{\rho}=-i[H,\rho]+2\sum_{j=1}^{N}\left[k_1\mathcal{D}[{a_j}^\dag](\rho)+k_2\mathcal{D}[{a_j}^2](\rho)\right],
\end{equation}
where, $H=\frac{V}{2d}\sum_{j=1}^{N}\sum_{m=j-d}^{m=j+d}\left({a_j}^\dagger a_m + a_j{a_m}^\dagger\right)$, $V$ is the coupling strength and $d$ is the coupling range with $m\neq j$ in the second sum.{\cc The quantum Stuart-Landau oscillator and quantum vdP oscillator have the same quantum master equation under the harmonic approximation; in the literature of quantum synchronization they are used interchangeably}. 
Here we consider the nonlocally coupled ring of quantum Stuart-Landau oscillators but in the presence of  Kerr nonlinearity; the modified Hamiltonian reads,
\begin{equation}
\label{kerr}
H=\sum_{j=1}^{N}K\left({a_j}^\dagger a_j\right)^2+\frac{V}{2d}\sum_{j=1}^{N}\sum_{m=j-d}^{m=j+d}\left({a_j}^\dagger a_m + a_j{a_m}^\dagger\right),
\end{equation}
where $K$ is the Kerr parameter which is identical for all the oscillators in the ring.

Eq.~\eqref{qchm_master} with the Hamiltonian given by Eq.~\eqref{kerr} is solved using the self-consistent method \cite{lee_prl} {\cc (see also \cite{qad1})}. As stated in Ref.~\cite{qchm}, we took the initial density matrix $\rho(t_0)$, which is the tensor product of coherent states centered around a specific group of $N$ points in the phase space. This group of points should obey the state of classical phase chimera, that means all these $N$ points will be at a fixed distance (equal to the amplitude of the classical oscillator) from the origin but a few consecutive points will be at the same phase (i.e., the angle between position vector and $x$-axis will be the same) while the phases of the rest of the consecutive oscillators will be random.

In the absence of Kerr nonlinearity, after a short time evolution the initial states will be evolved to squeezed states for each oscillator. The signature of quantum correlation appears in the form of the direction of squeezing \cite{qchm} {\cc(note that, the term squeezing is not used in its strict sense since the purity Tr$[\rho^2]$ of the phase-locking state of the quantum van der Pol oscillator is tiny and hence the system is mixed)}. Here we consider a nonlocally coupled ring network of $N=50$ quantum vdP oscillators without Kerr nonlinearity. Fig.~\ref{kerrchm}(a) represents the clear view of squeezed states of all the individual oscillators after the evolution of a short time interval $\Delta t=1$ from the initial state (we present the first 49 oscillators for visual presentation). It is clear that the direction of squeezing of $i\in(1,21)$ oscillators are identical (thus forming synchronous group) but the directions are random and disparate for the rest of the oscillators (thus forming the incoherent group). This is the quantum signature of chimera as discussed in Ref.~\cite{qchm}.

Next, we introduce Kerr nonlinearity in the system. Fig.~\ref{kerrchm}(b) represents the same scenario but in the presence of Kerr parameter $K=1$. The Kerr nonlinearity makes the squeezed state of both coherent and incoherent groups to ring shape oscillatory state thus making the phase of each oscillator indeterministic. In this way the distinction between the coherent and incoherent groups in the sense of the direction of squeezing is being destroyed and the symmetry breaking quantum chimera state converts to a symmetric oscillatory state. 
{\cc It is noteworthy that, from the perspective of phase-locking, a footprint of initial conditions remains in the phase locking states (although, higher values of $K$ remove this footprint). However, since in this paper, we follow the notion of \cite{qchm}, where the concept of phase-locking was not considered, rather, the notion of ``coherent" and ``incoherent" states were assigned according to the direction of squeezing in the phase space, our definition of symmetry does not depend on the phase-locking states. Also, barring chimera patterns, in general we can say that Kerr nonlinearity hinders phase-locking.} 
Unlike the QOD state, a strong measure of quantum chimeras is still elusive \cite{qchm}. Therefore, an exact value of $K$ where the transition from quantum chimera to symmetric state occurs could not be marked. However, from the qualitative evidence of the phase space structure we can infer that the Kerr nonlinearity is indeed detrimental to the quantum chimera state.

\section{Conclusions}
\label{con}
In this paper we have demonstrated that the Kerr anharmonicity is detrimental to symmetry breaking states in coupled quantum oscillators. We have considered two well known symmetry-breaking states, namely the quantum version of oscillation death and chimera state, and established that in both the cases the Kerr nonlinearity destroys those states and brings the coupled system to a rather symmetric configuration. Direct simulation of the quantum master equation and study with the semiclassical model have been employed to establish our findings. We further explored the effect of Kerr parameter on the quantum mechanical entanglement that is positively correlated with the symmetry-breaking steady state and showed that Kerr nonlinearity makes the entanglement weaker. 

{\cc We also studied all the scenarios for large Kerr parameter values, and got qualitatively the same results (not shown here). Since we considered identical oscillators, unlike  Refs.~\cite{qad2,brud-poch} we have not observed any strong quantum effects induced by the interplay of detuning and anharmonicity.} We believe that with the current technology, our results can be verified in experiments, e.g., using membrane-in-the- middle set up \cite{expt-mem} or ion-trap experiments \cite{lee_prl,expt-ion}. {\cc However, experimental realization of  conjugate coupling and a network of large number of oscillators will be challenging.} Further, quantum amplitude death was identified as a potential candidate for quantum mechanical cooling \cite{qad1}. Therefore, Kerr nonlinearity can be exploited for this purpose by inducing quantum amplitude death from quantum oscillation death state. Moreover, since symmetry-breaking steady state and the quantum entanglement are positively interconnected, our study suggests that the generation of entanglement can be engineered by controlling Kerr nonlinearity that will be useful in quantum information and computation applications \cite{entg1,qcomputation}.    

\begin{acknowledgments}
B.B. acknowledges the financial assistance from the University Grants Commission (UGC), India in the form of Senior Research Fellowship (SRF). T. B. acknowledges the financial support from the Science and Engineering Research Board (SERB), Government of India, in the form of a Core Research Grant [CRG/2019/002632].
\end{acknowledgments}

\providecommand{\noopsort}[1]{}\providecommand{\singleletter}[1]{#1}%
\end{document}